\documentclass[10pt,twocolumn]{article}

\usepackage{graphicx}
\usepackage{amsfonts}
\usepackage{amsmath}
\usepackage{amssymb}
\usepackage{mathrsfs} 
\usepackage{lipsum}
\usepackage{color}
\usepackage{authblk}
\usepackage[margin=2cm]{geometry}
\usepackage{mathtools, cuted}
\usepackage{bm}
\usepackage{amsmath}
\usepackage[utf8]{inputenc}
\usepackage[squaren,Gray,cdot]{SIunits}
\usepackage{caption}
\usepackage{subcaption}
\usepackage{mathtools, cuted}
\usepackage{array,amsmath}
\usepackage{textcomp}

\newcommand{\specialcell}[2][c]{%
\begin{tabular}[#1]{@{}c@{}}#2\end{tabular}
}

\begin{document}
\title{How does the presence of stevia glycosides impact surface bubbles stability?}
\author{Jonas Miguet$^1$, Yuan Fang $^3$, Florence Rouyer $^2$, Emmanuelle Rio$^1$\\
\small{$^1$ Univ. Paris Sud, Laboratoire de Physique des Solides, CNRS UMR 8502\\
$^2$ Laboratoire Navier, Université Paris-Est, 77454 Marne-la-Vallée, France\\
$^3$ PepsiCo Global R\&D, Valhalla, New York 10595, United States}
}

\maketitle 

\begin{abstract}
The addition of sweeteners in fizzy beverages not only affects the sugar content but also the bubbles stability. In this article, we propose a model experiment, in which the lifetime of hundreds of single bubbles is measured, to assess the stability of bubbles in solutions containing either sucrose or sweeteners. We show that the bubbles are indeed more stable in presence of sweeteners, which are surface active molecules and adsorb at the interface. Additionally, we test an antifoam at different concentrations and show that our experiment allows to identify the best concentration to reproduce the stability obtained in sucrose when we replace this latter by a sweetener. 
\end{abstract}

\section{Introduction}

The stability of bubbles at a liquid/air interface has been more and more explored in the past ten years \cite{Lhuissier2012, poulain2018,Poulain2018a,Champougny2016,kovcarkova2013film,Lorenceau2020} for the benefit of various applications ranging from the prediction of the climate  \cite{IPCC_AR5_CH7,murphy1998influence} to the food industry  \cite{Liger-Belair2009} or bacterial circulation \cite{Poulain2018a,blanchard1989ejection}.

The transfer of liquid from the oceans to the atmosphere and the production of condensation nuclei is indeed affected by the dispersion of aerosols by rupturing bubbles \cite{Johnson2013,Embil1997,Beerkens2006}.
On the other hand, similar aerosols are produced when bubbles rupture at the surface of carbonated beverages and contribute to the sensations of the drinkers through the dispersion of the different flavors.

\begin{figure}[h]
  \centering
  \begin{subfigure}[b]{\linewidth}
    \includegraphics[width=\linewidth]{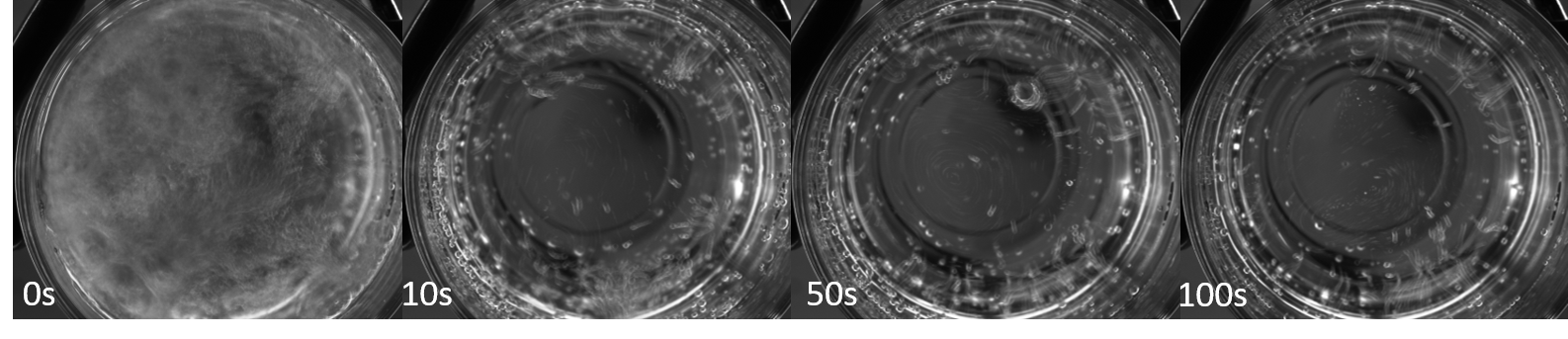}
    \caption{10$~wt\%$ sucrose solution}
    \label{fig:sucrose_pour}
  \end{subfigure}
  \vfill
  \begin{subfigure}[b]{\linewidth}
    \includegraphics[width=\linewidth]{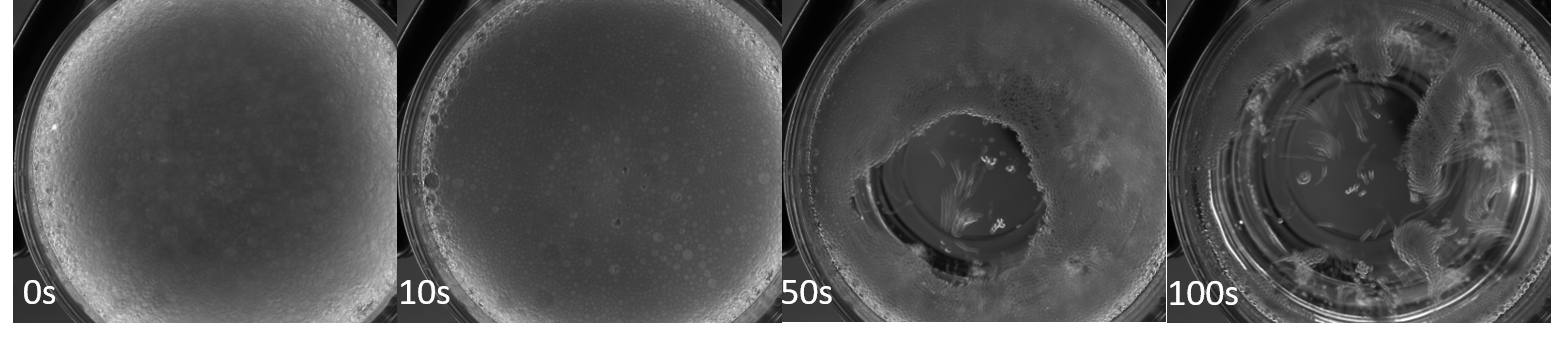}
    \caption{0.05$~wt\%$ sweetener solution}
	\label{fig:SG95_pour}  
  \end{subfigure}
  \caption{Foam evolution after the quick pouring of a solution in a glass, top view of a $\approx$ 7\~cm diameter glass filled containing around 15 cl. For the concentrations, the notation wt\% stands for a relative weight percent.}
  \label{fig:comp_sucrose_SG95}
\end{figure}

The starting point of the present work is the observation of the enormous difference between a foam produced by a sucrose-based carbonated beverage and a much more dilute sweetener-based one (see figure \ref{fig:comp_sucrose_SG95}). 
It turns out that the foam produced by the latter is more stable, therefore possibly altering the experience of the consumer through two mechanisms. 
First, a greater foamability and foam stability enhance the probability that the reckless pouring of the beverage 
leads to a high foam to liquid ratio, a longer waiting time before actual consumption of the liquid, or even to the overflowing of the glass.  
Second, as shown by Lhuissier \textit{et al} \cite{Lhuissier2012}, bubbles with a typical size beyond 1~mm are likely to produce so-called film aerosols upon bursting that are projected in the overlaying air above the glass and eventually deposited in gustatory sensors or evaporated, thus maximizing flavor sensations for the consumer \cite{Liger-Belair2009}. The number of these aerosols depends on the thickness of the film upon bursting, and therefore on their lifetime.

We propose here a set-up that allows to study single bubbles stability at the surface of a liquid at rest to compare quantitatively the lifetime of surface bubbles in beverages depending on their composition. 
Such physical systems  need to be studied statistically, since the lifetime of a bubble may inherently not be deterministic \cite{Vrij1966}. Nevertheless, it has been shown to be on average well-defined under controlled conditions \cite{saulnier2014study,Champougny2016,poulain2018}. We thus propose an experiment allowing to measure automatically the bubbles lifetime to extract statistically meaningful lifetime distributions in presence of sucrose and sweeteners. We also test the effect of antifoaming agents 
which are often added in industrial beverages to correct the enhanced foamability observed in presence of sweeteners.

\section{Material and Methods}

\subsection{Material}

\begin{figure}[h]
  \centering
  \begin{subfigure}[b]{0.3\linewidth}
    \includegraphics[width=\linewidth]{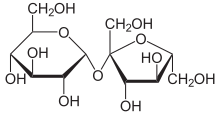}
    \caption{Sucrose}
    \label{fig:sucrose}
  \end{subfigure}
  \begin{subfigure}[b]{0.3\linewidth}
    \includegraphics[width=\linewidth]{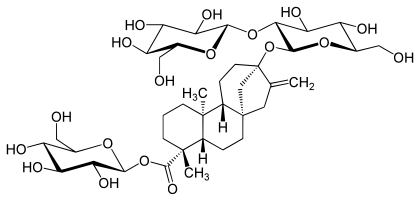}
    \caption{Stevioside}
	\label{fig:stevioside}  
  \end{subfigure}
  \begin{subfigure}[b]{0.3\linewidth}
    \includegraphics[width=\linewidth]{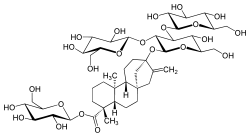}
    \caption{Rebaudioside A}
	\label{fig:rebA}  
  \end{subfigure}
  \caption{Cram Representations of sweetening molecules}
  \label{fig:Cram_representations}
\end{figure}

Different sweetening products have been tested and compared. 
Classical "white table sucrose" has been purchased from Sigma Aldrich ((S0389), GC grade, purity $\geq$ 99.5~\%).
The different sweeteners are natural extracts of \textit{stevia rebaudiana}: Stevioside and Rebaudioside A.
The three Cram representations of these molecules are pictured in figure \ref{fig:Cram_representations}. 
More precisely, we use REBA 97 and REBA 99, which are Rebaudioside A with a respective purity of 97 and 99 \%, the impurities being mostly other steviol glycosides.
We also tested SG95, which is the natural extract of \textit{stevia rebaudiana} purified up to 95 \% and thus a mixture of various steviol glycosides.
These three sweeteners are commercially available (PureCircles) and often used in industries.
As an antifoam, we used a food grade silicon antifoam (purchased from Momentive Performance Material Inc.). 

\subsection{Protocol for the solutions preparation}

All used concentrations are given in relative mass content [wt\%] \textit{ie}: $c_\text{product}=\frac{m_\text{product}}{m_\text{total}}$. 
The concentrations are chosen to reproduce the industrial concentrations in carbonated beverages. Since the purpose is to obtain a resulting taste as close as possible to that of sucrose but with sweeteners, the right parameter for this choice is the so-called relative sweetness that, in our case, is in the range 250-450 \cite{Brandle1998}(the manufacturer gives 230 and 270 respectively for stevioside and Rebaudioside A). Therefore, since the concentration of sucrose in cabonated beverages is of the order of $10~wt\%$, we took a concentration of $0.03~wt\%$ for all the sweetener-based ones.
All solutions are prepared using ultra clean water. 
We systematically add two preservatives to the solutions: Citric Acid and Sodium Benzoate at respective concentrations of $0.13~wt\%$ and $0.012~wt\%$.

All small masses (sweeteners, preservatives) are weighted using a precision scale (OHAUS Pionneer PA 214) and a weighting pan. Then the water is added and the full mass of the solution (or other big masses like for sucrose) is measured with a coarser scale (KERN 440-47N). Finally a micropipette is used to weight the antifoam, which is a liquid compound,
the concentration of which is assessed using a precision scale. For the solutions containing 0.1 or 1 ppm, the final concentrations are obtained by successive dilutions. The solutions are then stirred using a magnetic stirrer until complete dissolution.

Surface tension measurements have demonstrated the presence of impurities in sucrose (see SI). 
To ensure reproducibility of the experiments, we thus chose to filter all the solutions using a 0.22~\micro\meter ~wide pore-size cellulose membrane with the help of a Büchner system (funnel and flask) and a vacuum pump.
As shown in the Supplementary Information, this indeed allow to remove the impurities and to measure a constant surface tension during hours.
When some compounds should not be filtered (antifoams), the solutions is filtered prior to adding the corresponding products.

\subsection{Tensiometry}

To measure the surface tension of our solutions of interest, we used a commercial tracker (Teclis Instrument). The associated method is known as the "pendant drop" method although we rather use it in the "rising bubble" mode so that we avoid evaporation or depletion effects and minimize risks of pollution of the interface. The method consists in analyzing the shape of a bubble attached to a syringe 
\cite{Daerr}. All experiments are conducted at around 25~\degreecelsius. The results are given in Table \ref{table:sucrose_ST}. The surface tension is averaged between 10 and 35 s, which is the lifetime of the surface bubbles.

Another instrument has also been used that relies on a method referred to as Du Noüy-Padday. A small cylinder, with a diameter below 1 mm so that we can neglect buoyancy, is pulled out of the solution and the maximum pulling force $F_\text{max}$ [\newton] \textit{ie} the force at detachment is linked to the surface tension through:
\begin{equation}
	 \gamma=\frac{F_\text{max}}{2\pi r_\text{cylinder}}.
	 \label{FtoST}
\end{equation}
where $\gamma$ [\newton\per\meter] is the surface tension and $r_\text{cylinder}$ [\meter] the radius of the cylinder. Measurements are repeated 10 times and results are given together with the measured standard deviation in Table \ref{table:sucrose_ST}.

We show in Appendix B that both methods give comparable results.

\begin{table*}
\begin{center}
\begin{tabular}{ |c|c|c|c| }
 \hline
  Solution & Method &  Mean [mN/m] &  \specialcell{ Standard \\ Deviation [mN/m]}\\
 \hline
\specialcell{Water} & Du Noüy-Padday & 72.05  &  0.06 \\
\specialcell{Sucrose [10 wt\%] \\ (without preservatives)} & Du Noüy-Padday & 72.30 & 0.5 \\
\specialcell{Reba 97} & Tracker & 62.7 & 0.3 \\
\specialcell{Reba 99} & Tracker & 62.2 & 0.2 \\
\specialcell{SG 95} & Tracker & 62.4 & 0.3 \\
\specialcell{Reba 97 + Antifoam [0.01 w\%]} & Tracker & 63.6 & 0.2 \\

\hline 
\end{tabular}
\caption{Surface Tension of the different solutions.}
\label{table:sucrose_ST}
\end{center}
\end{table*}

\subsection{Evaporation}

We measured the evaporation rate for different solutions in order to check whether this parameter is critical or not. To do so, we pour the solution of interest up to the surface of a petri dish of diameter 98.63~mm. We then place it on a high precision scale in a closed chamber in which we can control the relative humidity. We set the humidity to 50 \% and wait for at least half an hour for the regulation to proceed. 

\begin{figure}[!ht]
	\centering
		\includegraphics[width=\linewidth]{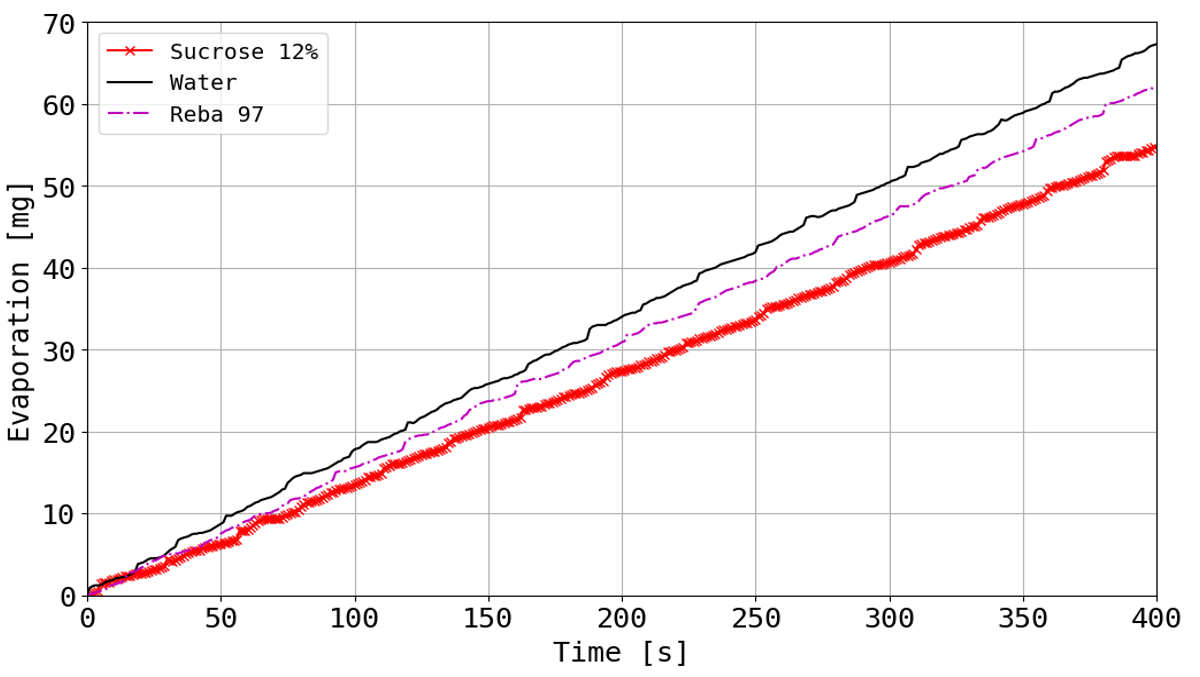}
		\caption{Mass contained in a petri dish along time while evaporation occurs in an atmosphere controlled at 50 \% humidity.}	\label{fig:evaporation}
\end{figure}

The weight of the petri dish $m(t)$  [\milli\gram] is measured along time (Figure \ref{fig:evaporation}) and the mass evaporation rate $F$ [\milli\gram\per\second] is directly the slope of the mass loss versus time data obtained by linear regression (Table \ref{table:evaporation}).

\begin{table}[h]
\begin{center}
\begin{tabular}{ |c|c| }
 \hline 
 Solution & Evaporation rate F [mg/s] \\
 \hline
Water & 0.169\\
Sucrose & 0.137\\
Reba 97 & 0.155\\
 \hline
 \end{tabular}
\caption{Evaporation rates obtained from the fitting of Figure \ref{fig:evaporation} for each solution.}
\label{table:evaporation}
\end{center}
\end{table}

\subsection{Experimental Set-Up for bubble lifetime measurement}

\begin{figure}[!ht]
  \centering
  \begin{subfigure}[b]{0.9\linewidth}
    \includegraphics[width=\linewidth]{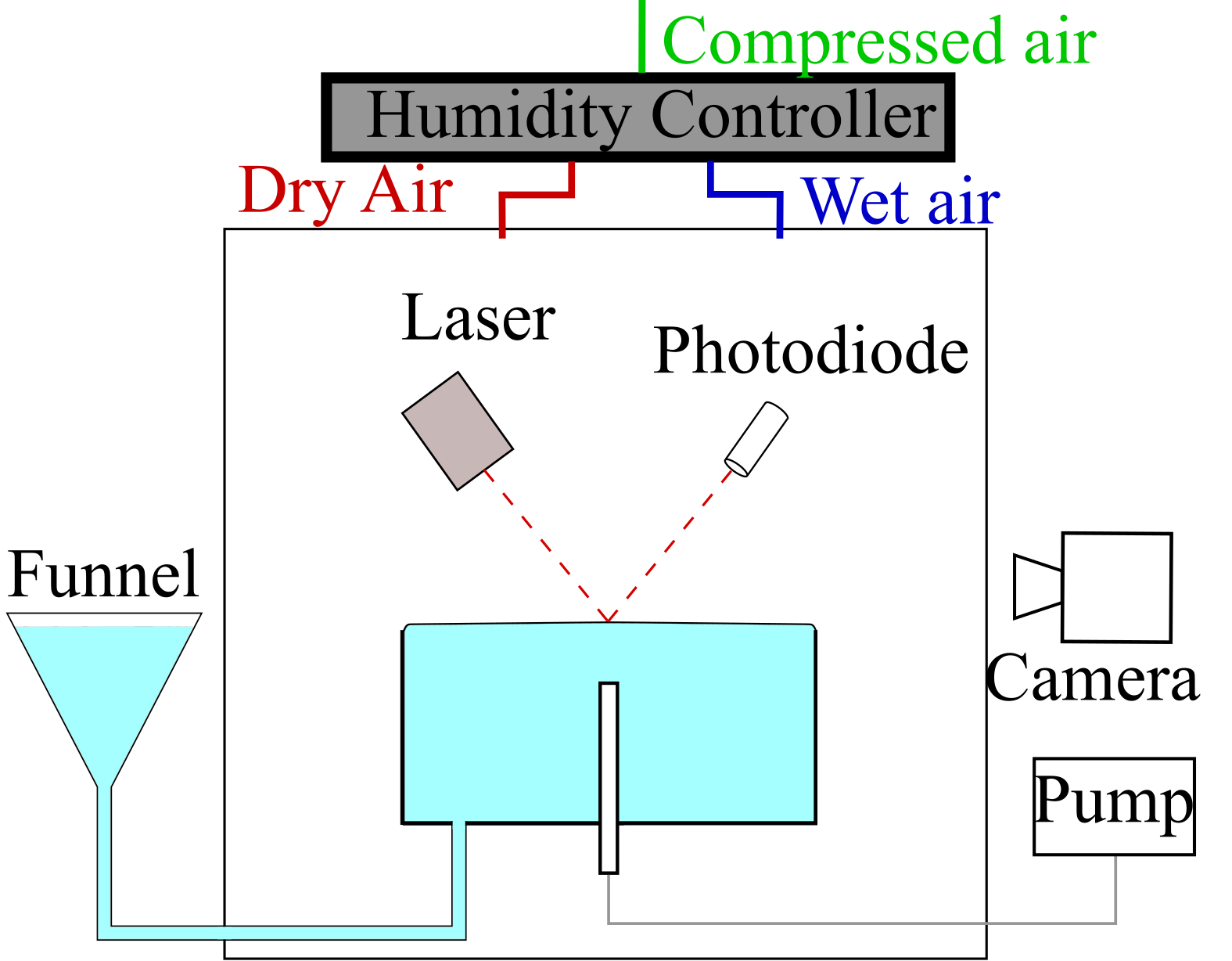}
    \caption{Undisturbed interface, the laser beam is reflected into the photodiode}
    \label{fig:undisturbed}
  \end{subfigure}
  \begin{subfigure}[b]{0.9\linewidth}
    \includegraphics[width=\linewidth]{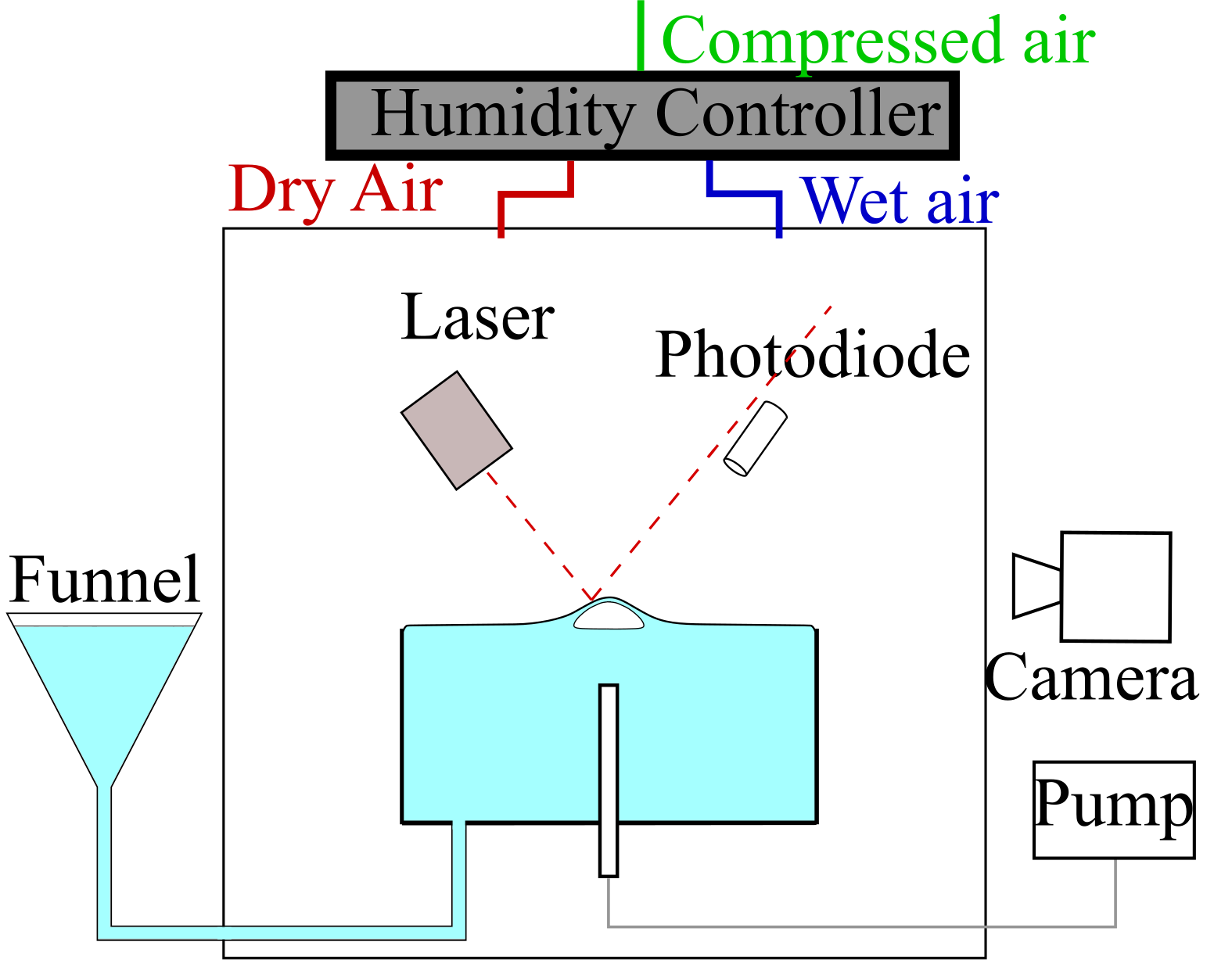}
    \caption{As a bubble rises to the surface, the laser beam is reflected away from the detector}
	\label{fig:bubble}  
  \end{subfigure}
  \caption{In a closed humidity controlled chamber, with a humidity fixed at 50 \%, a container is filled through a funnel. Air is injected at the bottom of the container by a pump to create bubbles, the presence of which is assessed using a laser and a photodiode. Images are recorded from the side during the experiments.}
  \label{fig:chronobulle}
\end{figure}

Figure \ref{fig:chronobulle} shows the set-up that allowed us to repeat measurements of the bubbles lifetime under controlled conditions. The solution of interest is put in a cylindrical container of 4~cm of inner diameter. The liquid/air surface is positioned slightly above that of the container, taking advantage of the meniscus, so that the bubbles can be imaged from the side. At the bottom of the cylinder, a hole is pierced and sealed with a piece of rubber. A stainless steel needle is then vertically planted in the rubber.
The syringe is placed so that the drop is generated at a distance of 1 cm from the surface. .This implies, following the results of Zawala et al\cite{Zawala2011}, and given our bubble size, that the bubble velocity is close to its free ascension velocity and therefore closer to the application conditions.
It is dressed with a PTFE (polytetrafluoroethylene, hydrophobic) tube of inner diameter 0.4~mm and 1~cm long to avoid any wetting of the injector. The bottom of the needle, outside the cylinder is plugged into a series of tubes going through electronically controlled valves that drive the injection of bubbles by a flow controlled aquarium air pump.

A laser beam is directed to the surface of the  liquid bath, where it is directly reflected into a photodiode when the interface is at rest (Figure \ref{fig:undisturbed}) or diverges when a bubble is present (Figure \ref{fig:bubble}). 
This automated assessment of the presence of a bubble, allows for repeated measurements by creating a new bubble 10~s after the detection of the rupture of the previous one. 
We are aware that, when a bubble reaches the interface it is susceptible to bounce. 
However, we never observed this phenomena probably because it is too fast. Indeed, following the results of Zawala et al\cite{Zawala2011}, we can estimate that this phenomenon lasts up to hundreds of milliseconds, which is of the order of our measurements resolution (the frame rate of the camera is 3.75 s$^{-1}$) and can safely be neglected in the development that we propose.
We conducted similar experiments, only varying the waiting time between two successive bubbles to check its impact and did not notice any sensible change. 

\begin{figure}[!ht]
	\centering
		\includegraphics[width=\linewidth]{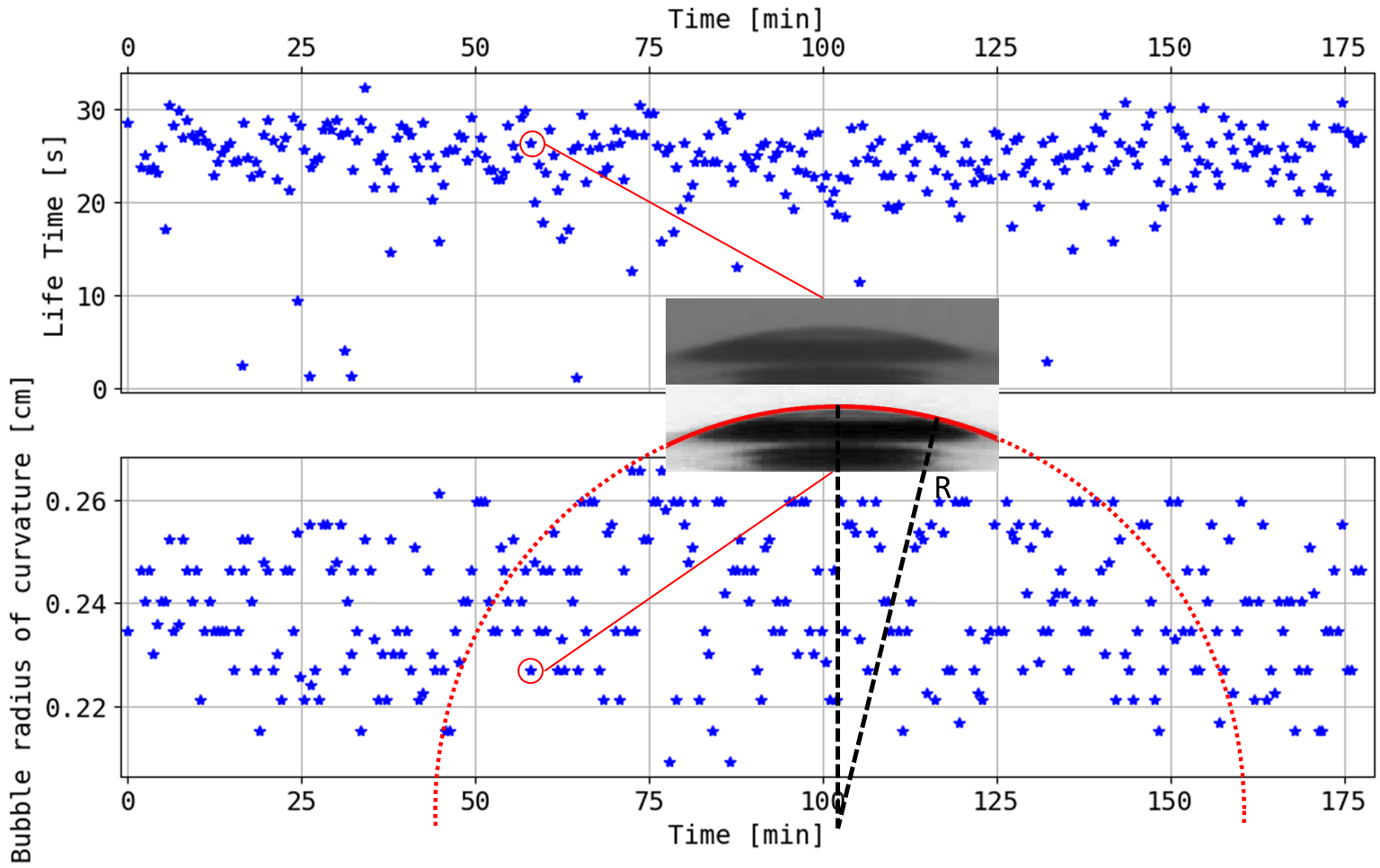}
		\caption{Example of raw data set for Reba 97. The relative humidity is regulated at 50\%. Each point corresponds to a new bubble with a given lifetime and a given bubble radius. (Top) Lifetime of the bubbles as a function of the experimental time. (Bottom) Size of the bubbles as measured by image analysis, as a function of the experimental time. The pictures represent an example of the same raw (top) and processed (bottom) image together with the extracted radius of curvature of the spherical cap $R$.}	\label{fig:raw_dataset}
\end{figure}

The continuous capture of the side images provides both the size of the bubbles and a reliable measure of the lifetime.
This direct measurement allows to avoid the artifacts inherent to the laser detection, \textit{i.e} presence of very stable daughter bubbles \cite{Bird2010} or loss of the laser alignment due to evaporation.
The image processing is done using the scikit-image python library to get the lifetime $\pm$0.4 s of the bubbles. As soon as a bubble is detected, its size is measured with a resolution of 30$\mu$m, making use of a fitting algorithm based on Hough transform.
The full reconstruction of a given serie of measures, making use of the time given by the electronic device connected to the photodiode and that of the camera, allows to ensure that no bubble is missed and that the lifetime is always precisely measured within one image for the birth and the death of a given bubble. The frame rate of the camera being 3.75 s$^{-1}$, our resolution is the limiting factor for the measure which uncertainty is then estimated to be 0.4 s. Figure \ref{fig:raw_dataset} shows the lifetime of individual bubbles (top chart) and their measured size (bottom) chart, as a function of the time elapsed since the beginning of the experiment. This typical experiment has been obtained using Reba 97. The size of the bubbles is very reproducible with an average bubble cap radius of 0.24 cm and a standard deviation of 0.013 cm. This parameter  will be kept constant throughout the experiments. The lifetime of individual bubbles does not follow any particular trend during the whole experiment. This is a sign that the system is stable along time. 

To clean the tank between each experiment, the system can be rinsed without having to open the box, in which the atmospheric humidity is controlled. 
Three rinses are made with ultrapure water and one with the next solution of interest, before filling the tank. 
After using anti-foaming products that are oil-in-water emulsions, we need to open the chamber, clean the funnel assembly of the tank hose with washing-up liquid, rinse thoroughly and re-establish the humidity control in the chamber before repeating the measurement.

Finally, we checked the property of the entire set-up before each experiment by testing the lifetime of surface bubbles in pure water. The surface bubbles then burst a second or less after their generation indicating that the pure water has not been polluted.

\section{Results and Discussion}

All the results for the bubbles stability are presented in the form of lifetime distributions (Figures \ref{fig:compare_sugars}, \ref{fig:compare_antifoams}). The number of bubbles $n$ is written in the legend. The y-axis represents the normalized probability density function (labelled PDF). The number of bins $N_\text{bins}$ is 100. The integral of the PDF being equals to 1 by definition, this condition leads, in our discrete case, to $\sum_{i=1}^{N_\text{bins}}N_\text{i}dt=1$ with $N_\text{i}$ the normalised probability to get a given value of the lifetime within a given interval and $dt$ the width of the intervals. $dt$ is constant and equals 0.5 s. The radii along with their standard deviations are summarized in table \ref{table:all_results} together with the parameters obtained by fitting the distributions as explained in the following.

\begin{table*}[h]
\begin{center}
\begin{tabular}{ |c|c|c|c|c|c| }
 \hline 
 Solution & n & \specialcell{Mean \\ Radius [mm]} & \specialcell{Std \\ Radius [mm]} & \specialcell{$\mu$ [s]} & \specialcell{Std \\ Lifetime [s]} \\
 \hline
 Sucrose  & 575 & 2.89 & 0.17 & 13.60 & not applicable \\  
 Reba97 & 305 & 2.40 & 0.13 & 25.90 & 2.73 \\
 Reba99 & 338 & 2.51 & 0.014 & 23.96 & 2.42 \\
 SG95 & 199 & 2.24 & 0.015 & 25.30 & 2.60 \\
 Reba97 + antifoam 0.1 ppm & 306 & 2.33 & 0.06 & 8.69 & 4.31 \\
 Reba97 + antifoam 1 ppm & 313 & 2.13 & 0.01 & 2.34 & 0.85 \\
 Reba97 + antifoam 50 ppm & 103 & 2.51 & 0.15 & 1.32 & 0.29 \\
\hline
 \end{tabular}
\caption{Recap chart of all lifetime experiments}
\label{table:all_results}
\end{center}
\end{table*}

\subsection{Comparison of bubbles stability for different sweetening products}

\begin{figure}[h]
	\centering
		\includegraphics[width=\linewidth]{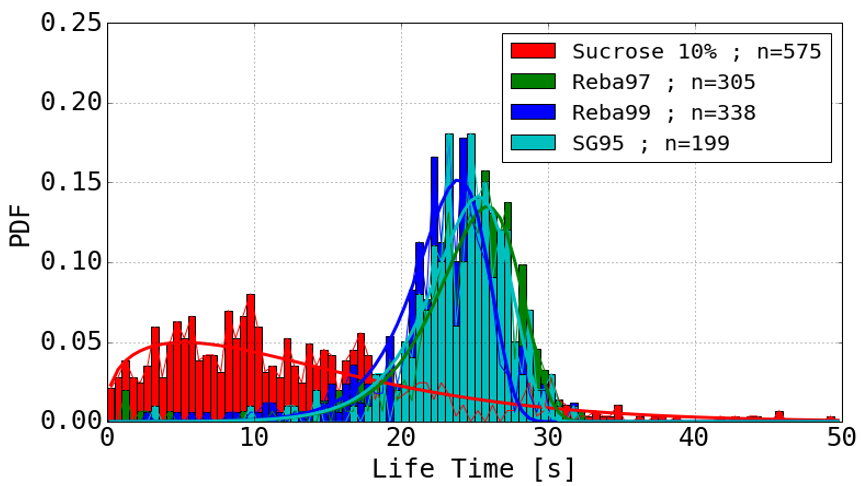}
		\caption{Lifetime distributions of the different tested sweetening products. $n$ in the legend represents the number of bubbles for a given distribution. The relative humidity is regulated at 50\%.}
	\label{fig:compare_sugars}
\end{figure}

In this section, we will compare the effect of the different sweeteners and of the sucrose on the stability of the bubbles. Figure \ref{fig:compare_sugars} represents the lifetime distributions for bubbles created with the different products. 

The first result is that the presence of sucrose stabilizes the surface bubbles, which live longer than in pure water. Indeed, as mentioned in the experimental set-up, the stability of bubbles made in pure water vanishes because no stabilizing effect exist in absence of surfactants which usually generate a Marangoni elasticity that can sustain the weight of the cap. 

A comparison between the surface tension of water and that of 10 wt\% sucrose solutions is summarized in table \ref{table:sucrose_ST}. The surface tension is only very little affected with 10 wt\% sucrose as compared to water which is consistent with previous studies and the model of Docoslis et al. \cite{Docoslis2000} including a depletion zone close to the interface.
Additional differences between the sucrose and the water solutions are the viscosity and the evaporation rate, which can contribute to the drainage dynamics of the film and therefore affect its stability. The bulk viscosity of a sucrose solution of concentration 10~wt\% is approximately 25~\% higher than that of water at 20~\degreecelsius \cite{Telis2007}. Second, as shown in figure \ref{fig:evaporation} and table \ref{table:evaporation}, the addition of sucrose decreases the evaporation rate of the solution. Both pledge for a decreased thinning rate of the film.

A model has been proposed recently by Poulain \textit{et al} \cite{Poulain2018a} to give an estimation of the bubble lifetime in presence of surfactants. We proposed a modified model \cite{Miguet2020} by including a convective evaporation. 
This model includes the influence of the surface tension $ \gamma$, of the viscosity $\eta$ and of the evaporation rate $J$.
The main physical ingredient is the comparison between the thinning rate due to drainage and the one due to evaporation. 
If $h_c$ is the thickness for which both thinning rates are equal, the lifetime $\tau$ is given by $\frac{h_c}{J}$.
This model leads to a bubble lifetime 
\begin{equation}
\tau \simeq \frac{R^{7/5} \eta^{2/5}\rho^{3/5}}{J^{3/5}\gamma^{2/5}\ell_c^{2/5}}
\label{eq:tau}
\end{equation} 
 where $\ell_c=\sqrt{\frac{\gamma}{\rho g}}$ is the capillary length depending on the density $\rho$ of the liquid and on the gravitational acceleration $g$.

We calculated $\tau$ for the different sweeteners using the surface tension in table \ref{table:sucrose_ST}, a density of $1000$ kg/m$^3$ and the radius in table \ref{table:all_results}. The viscosity is $\eta=10^{-3}$ Pa.s for every solution but sucrose for which we took $\eta=1.26 \times 10^{-3}$ Pa.s.
The evaporation rate $J$ is estimated by using the evaporation rate $F$ given in table \ref{table:evaporation} as $J=\frac{F}{F_{\text{water}} }\times J_{\text{water}}$, with $J_{\text{water}}=3.64\times10^{-5}$ kg/m$^2$/s \cite{Boulogne2018}. 
The underlying hypothesis is that the local evaporation rate $J$ scales as the measured global evaporation rate $F$.
The calculated values of $\tau$ are reported together with the experimental value of the bubbles lifetime in Table \ref{table:tau}.
This scaling appears to overestimate the lifetime of surface bubbles in sucrose by a factor of almost 2.

\begin{table}[h]
\begin{center}
\begin{tabular}{ |c|c|c|}
 \hline 
 Solution & Experimental lifetime [s] & $\tau$ [s]\\
 \hline
 Sucrose  & 12.5 & 24.1 \\  
 Reba97 & 25.9 & 16.6\\
 \hline
 \end{tabular}
\caption{Comparison of the experimental lifetime with the value of $\tau$.}
\label{table:tau}
\end{center}
\end{table}

The second important result here is that all the sweeteners have the same quantitative effect on the stability of bubbles. Indeed, the stability of the bubbles is enhanced in all cases for sweeteners as compared to sucrose, which is consistent with the first remark we made about carbonated beverages made of sweeteners featuring more stable foams (Figure \ref{fig:comp_sucrose_SG95}). Moreover, it appears very clearly that no significant difference regarding bubbles stability can be found among the different sweeteners. 
Table \ref{table:sucrose_ST} summarizes the results of surface tension measurements using the tracker for each sweetener. 
No noticeable effect of the proportion of the blend or the nature of the molecules used can be seen here. 
The decreased surface tension of these products as compared to pure water is a sign that they have an affinity for the interface. 
This is qualitatively explained by the structure of these molecules (Figure \ref{fig:Cram_representations}). Stevioside and Rebaudioside A indeed feature both a hydrophobic part (the aromatic compounds) and some hydrophilic parts (the other aliphatic chains surrounded by hydroxyl groups that can make hydrogen bonds with water). 
Moreover, as shown by the value calculated in Table \ref{table:tau}, the model proposed by Poulain \textit{et al} gives values very close to the experimental lifetime. Indeed, these amphiphilic groups can reach the interface so that we expect a behavior closer to the one of surface bubbles stabilized by surfactants.

The third result is that not only the average value but also the lifetime distribution is very different for sweeteners and for sucrose. 
In Figure \ref{fig:compare_sugars}, the distributions obtained with sweeteners are indeed fitted by an extreme value distribution whereas the curve obtained for sucrose is fitted by a Weibull probability density function.

The extreme value probability density function has two parameters, $\mu$, the location parameter, which indicates the position of the maximum and $\sigma$, a scale parameter indicating the width of the distribution:
\begin{equation}
    E(t) = \frac{1}{\sigma}\exp{\frac{t-\mu}{\sigma}}\exp{\left(-\exp{\frac{t-\mu}{\sigma}}\right)}.
    \label{eq:ExtremeValue}
\end{equation}
This distribution exhibit an asymmetric bursting probability.
In our case, an interpretation is the presence of a well-defined characteristic time located around $\mu$ together with early bursting accidents, i.e. bubbles bursting at earlier time.

The Weibull function has been proposed by Lhuissier \textit{et al} in \cite{Gilet2007,Lhuissier2012} to describe the bursting time distribution in dirty water. It contains one free parameter $\mu$ linked to the position of the maximum:
\begin{equation}
    W(t) = \frac{4}{3}\frac{t^{1/3}}{(0.92\mu)^{4/3}}\exp{-\left(\frac{t}{0.92\mu}\right)^{4/3}},
    \label{eq:Weibul}
\end{equation}
and well describes the data obtained in sucrose because of the long tail.
This suggests a dominating stochastic behavior for these systems and therefore, in average, a longer time for a hole to nucleate than for the film to thin down. 

Finally, our interpretation is that the mechanism leading to bubble bursting in sucrose is closer to the one observed in dirty water, whereas the physical picture in the presence of sweeteners seems to be qualitatively different and closer to surfactant based systems. The higher surface tension of sucrose as compared to sweetener-based systems indeed pledges in favour of a mechanism similar to that proposed by Poulain \cite{poulain2018} for dirty water. In this mechanism, a soluble impurity adsorbs to both sides of the film, locally reduces the surface tension and leads to a divergent Marangoni flow, a subsequent local thinning and the final rupture. This mechanism should lead to bursting events for a larger film thickness. On the other hand, the good correspondence of the sweetener solutions with equation 2 seems to be qualitatively in agreement with surfactant-based systems for which the lifetime is ultimately governed by evaporation \cite{Poulain2018a,Miguet2020}.

This first experiment demonstrates that our experimental apparatus gives the right diagnostic concerning the prediction of foamability in a gazeous beverage. 
The sweeteners are indeed molecules, which can populate the interface and account for additional Marangoni stresses able to counterbalance the weight of the liquid film and to stabilize the surface bubbles as well as the foams.
In the following, we will use our diagnostic at the scale of surface bubbles to test if antifoaming agents can destabilize surface bubbles so that a lifetime distribution similar to the one observed for sucrose can be achieved.

\section{Bubbles stability in presence of an antifoam}

\begin{figure}[h]
	\centering
		\includegraphics[width=\linewidth]{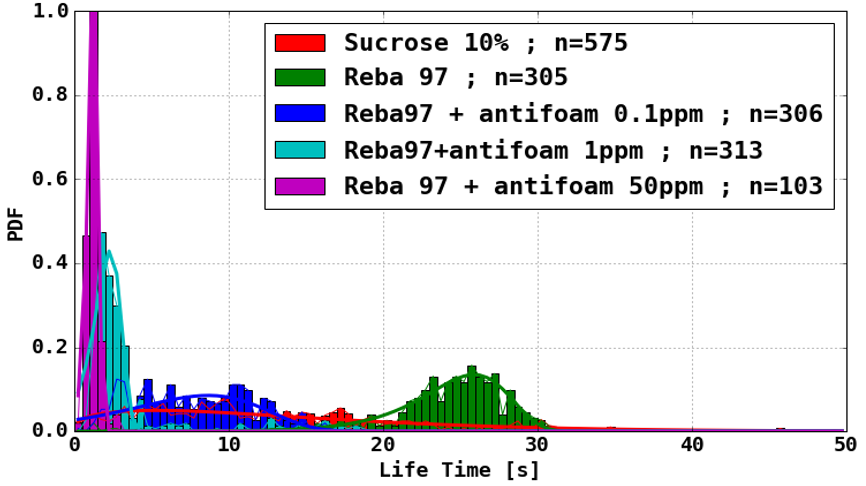}
		\caption{Lifetime distributions of Reba97 and the antifoaming agents. "n" in the legend represents the number of bubbles for a given distribution. The relative humidity is regulated at 50\%.}
	\label{fig:compare_antifoams}
\end{figure}

To measure the efficiency of an antifoaming agent on the stability of sweetener solutions, we select  Reba 97 as a sweetener.
The results in presence of the antifoam at different concentrations are represented in figure \ref{fig:compare_antifoams}. At high concentration, the antifoam unsurprisingly completely killed the stability of the bubbles. No difference can be measured in our system between these solutions and ultrapure water. The highest possible efficiency of the antifoaming properties of these products is therefore assessed in this system. In the case of the antifoam at smaller concentrations, the obtained distribution is intermediate and we identified that a concentration of 10$^{-4}$~wt\% (0.1 ppm) allows to recover an average lifetime close to the one observed with sucrose. Nevertheless, the distribution is still better described by an extreme value distribution and not by a Weibull function.
This makes this antifoam at this concentration a good candidate to be added to fizzy beverages in presence of sweeteners to recover a bubble stability close to the one observed in presence of sucrose.

\section{Conclusion}

The impact of the physical chemistry of different chemical components potentially found in sodas on the stability of single bubbles has been measured. The stability is assessed by measuring the lifetime of the bubbles. The temperature and the humidity of the environment are controlled, as these parameters are known to affect the system. 

The results confirm that single bubbles made with sweeteners, despite a product concentration a few hundred times smaller as compared to sucrose, are significantly more stable. 
No stability difference was found among the tested sweeteners. 
This confirms the empirical observation of enhanced foamability and foam stability of sweetener-based sodas as compared to sucrose-based ones. 
Additionally, our work demonstrate that the replacement of the sucrose by sweeteners also affects the lifetime distribution.
Different antifoaming agent concentrations were tested along with the same sweetener, Reba 97 and we showed that an intermediate concentration allows to reproduce the average lifetime of bubbles in a sucrose solution. 

These result show how any variation of the recipe can affect the bubbles and foam stability in fizzy beverages, which in turn, can affect the consumer experience. Our experiment suggests that the stability of surface bubbles is a simple controlled system, which can help formulating fizzy beverages. Indeed, since the atmospheric  conditions as well as the bubble size are well controlled, our conclusions really concern the formulation. 

We would like to emphasize that such a study is only a first step towards a comprehensive understanding of the effect of sweeteners on carbonated beverages. 
For example, here we consider only air as a gas for obvious experimental reasons whereas, in carbonated beverages, the gas contained in the bubble is C0$_2$ \cite{Karakashev2009,Weissenborn1996}. 
A validation of the results for a different gas is beyond the scope of this article but would be very interesting to validate the applicability of our conclusions in real fizzy drinks.

\section{Acknowledgments}
We acknowledge the technical contribution of Christophe Courrier for the electronic interface, David Hautemayou and Cédric Mézière for ensuring the mechanical support during the design of the experiment, Vincent Klein and Jérémie Sanchez for building the humidity controller and Laura Wallon for surface tension measurements. This work was funded by PepsiCo R\&D. The views expressed in this manuscript are those of the authors and do not necessarily reflect the position or policy of PepsiCo Inc.

\section*{Appendix A}
We compared surface tension measurements for a high quality sucrose (S0389, GC grade, purity $\geq$ 99.5~\%) before and after filtration. The data were obtained using the set-up with the cylinder in the dynamic surface tension mode described in the material and methods section. The surface tension decreasing with time for unfiltered sucrose is the sign of the presence of a pollution in the solution. On the other hand the constant surface tension observed in sucrose after filtration is a signature that the sucrose is pure enough for our purpose. We thus decided to filter every solutions.

\begin{figure}[h]
	\centering
		\includegraphics[width=\linewidth]{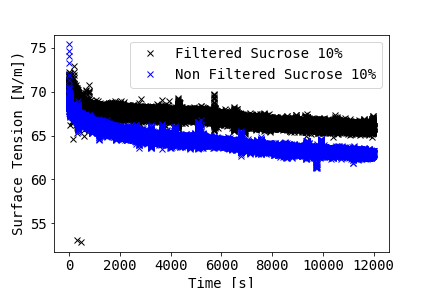}
		\caption{Dynamic surface tension of 20~wt\% sucrose solutions after and before filtration. The surface tension was measured by the Du-Noüy-Padday method.}
	\label{fig:ST_sucrose}
\end{figure}

\section*{Appendix B}
\begin{figure}[h]
	\centering
		\includegraphics[width=\linewidth]{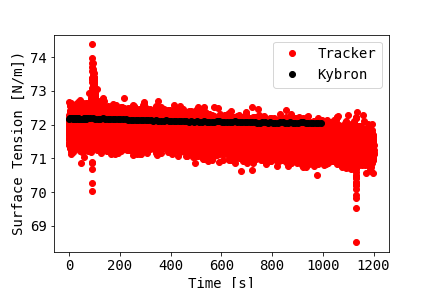}
		\caption{Measurement of surface tension of pure water with both measurements methods.}
	\label{fig:CompareKybronTracker}
\end{figure}

We have used to very classical commercial setups to measure the surface tension.
The first one, which is called the Tracker uses the rising bubble method.
The second one, which we call the Kybron uses the du Noüy-Padday method.
We measured pure water with both methods to show that they both give very similar results (Fig. \ref{fig:CompareKybronTracker}).

\bibliographystyle{unsrt}
\bibliography{library}
\end{document}